\documentclass{ws-ijmpd}
\begin{document}

\markboth{Tahereh Azizi, Emad Yaraie} {Cosmological Dynamics of
Modified Gravity With a Non-minimal Curvature-matter Coupling}

%
%

\title{COSMOLOGICAL DYNAMICS OF MODIFIED GRAVITY WITH A NON-MINIMAL CURVATURE-MATTER COUPLING}

\author{TAHEREH AZIZI}

\address{Department of Physics, Faculty of Basic
Sciences, University of Mazandaran,\\
Bobolsar 47416-95447, Iran\\
t.azizi@umz.ac.ir}

\author{EMAD YARAIE}

\address{Department of Physics, Faculty of Basic
Sciences, University of Mazandaran,\\
Bobolsar 47416-95447, Iran\\
e.yaraie@stu.umz.ac.ir}

\maketitle

\begin{history}
\received{Day Month Year}
\revised{Day Month Year}
\end{history}

\begin{abstract}
We perform a phase space analysis of a non-minimally coupled
 modified gravity theory with the Lagrangian density of the form $\frac{1}{2} f_{1}(R) +[1+\lambda
f_{2}(R)]{{\cal{L}}_{m}}$, where $f_1(R)$ and $f_2(R)$ are arbitrary
functions of the curvature scalar $R$ and ${{\cal{L}}_{m}}$ is the
matter Lagrangian density. We apply the dynamical system approach to
this scenario in two particular models. In the first model we assume
$f_1(R)=2R$ with a general form for $f_2(R)$ and set favorable
values for effective equation of state parameter which is related to
the several epochs of the cosmic evolution and study the critical
points and their stability in each cosmic eras. In the second case,
we allow the $f_1(R)$ to be an arbitrary function of $R$ and set
$f_2(R)=2R$. We find the late time attractor solution for the model
and show that this model has a late time accelerating epoch and an
acceptable matter era.
\end{abstract}
 \ccode{PACS numbers: 04.50.Kd, 95.36.+x, 98.80.-k,
98.80.Jk}


\section{Introduction}
Recent observational data suggest that our universe is undergoing an
accelerating phase of expansion
[\refcite{Riess98}-\refcite{Riess04}]. One way to explain the cosmic
speedup within the framework of general relativity is adding a
mysterious component to the matter content of the universe which is
dubbed as dark energy. Several candidates to the dark energy are
proposed such as cosmological constant, the scalar fields, Chaplygin
gas and so on (see [\refcite{Amendola10}] and references therein).
Another popular approach to describe the accelerating expansion of
the universe is to assume a modification to the general theory of
relativity on cosmological scales. In this respect the so-called
$f(R)$ modified gravity theory has attracted a lot of attention
[\refcite{Capozziello03}-\refcite{Nozari09}]. In these theories, the
curvature scalar $R$ is replaced by a generic function of $R$ in the
action. For a review of $f(R)$ modified gravity and its cosmological
implications see [\refcite{Sotiriou10}-\refcite{Nojiri11}].
Recently, an extension of $f(R)$ theories has been proposed in
[\refcite{Bertolami08a}] which there is an explicit coupling between
matter Lagrangian density and the curvature scalar (see also
[\refcite{Nojiri04}] and [\refcite{Allemandi05}]). This model leads
to considerable cosmological implications which some of them are:
the energy exchange between the matter fields and the curvature and
deviation from geodesic motion [\refcite{Bertolami08a}], the
mimicking of dark matter by leading to the flattening of the galaxy
rotation curves[\refcite{Bertolami10a,Bertolami12}], the modelling
of the cosmic speed up at late times [\refcite{Bertolami10b}] and
the reheating scenario after inflation [\refcite{Bertolami11}]. The
cosmological perturbations and the matter density perturbations of
this scenario are studied in [\refcite{Nesseris09}] and
[\refcite{Bertolami13}] respectively.

The aim of this paper is to study the dynamics of the modified
gravity theory with a non-minimal coupling between matter and
geometry via a phase space analysis approach. The strategy is to
rewrite Einstein's field equations for cosmological models in terms
of an autonomous system of ordinary differential equations
(ODE)[\refcite{Carloni05}-\refcite{Nesseris07}]. The paper is
organized as follows. In Section 2 we introduce the non-minimal
modified gravity scenario and its gravitational field equations.
Using the flat Friedman-Robertson-Walker metric with
 a perfect fluid form of the stress energy tensor, we obtain the generalized Friedmann equations
 of the scenario. In order to study the cosmological dynamics of this model, we consider two particular
cases in the rest of this paper. In section 3 we set $f_1(R)=2R$ and
a general form for $f_2(R)$
 and apply a phase-space analysis approach to obtain the critical points and their stability
in several eras of the cosmic evolution. Setting $f_2(R)=2R$ and a
power low form for $f_1(R)$, in section 4, we study the dynamical
system and analyze the nature of the resulted critical points of
this system. Finally, section 5 is devoted to conclusions.
\section{The Equations of Motion}
The action of the modified gravity model with a non-minimal coupling
with matter is given by [\refcite{Bertolami08a}]
\begin{equation}
S=\mathop{\int }\Big[\frac{1}{2} f_{1}(R) +[1+\lambda
f_{2}(R)]{{\cal{L}}_{m}} \Big]\sqrt{-g} d^{4}{x}\, .
\end{equation}
where ${{\cal{L}}_{m}}$ is the matter Lagrangian density, $f_1(R)$
and $f_2(R)$ are arbitrary functions of the curvature scalar $R$ and
$\lambda$ is a coupling parameter. Varying the action (1) with
respect to the metric $g_{\mu\nu}$ yields the following
gravitational field equations [\refcite{Bertolami08a}]
$$
[f'_1(R)+2\lambda
f'_2(R){{\cal{L}}_{m}}]R_{\mu\nu}-\frac{1}{2}g_{\mu\nu}f_1(R)=
$$
\begin{equation}
(\nabla_{\mu}\nabla_{\nu}-g_{\mu\nu}\Box)(f'_1(R)+2\lambda
f'_2(R){{\cal{L}}_{m}})+[1+\lambda f_2(R)]T_{\mu\nu},
\end{equation}
where a prime denotes the derivative with respect to the curvature
scalar. $T_{\mu\nu}$ is the stress-energy tensor and is related to
the Lagrangian density of the matter as follows
\begin{equation}
T_{\mu \nu }=-\frac{2}{\sqrt{-g}} \frac{\delta \left(
\sqrt{-g}{\cal{L}}_{m}\right) }{\delta g^{\mu \nu }}\,.
\end{equation}
Taking the trace of Eq.(2) leads us to
\begin{equation}
[f'_1(R)+ 2\lambda f'_2(R){{\cal{L}}_{m}}]R- 2f_1(R)=-3\Box[f'_1(R)
+ 2 \lambda f'_2(R) {{\cal{L}}_{m}}] +[1+\lambda f_2(R)]T
\end{equation}
where $T$ is the trace of the stress-energy tensor. We assume that
the matter content of the universe is described by a perfect fluid
with an energy momentum tensor
\begin{equation}
T_{\mu\nu} = (\rho + p ) u_{\mu} u_{\nu}+ p g_{\mu\nu},
\end{equation}
where $u^{\mu}$ is the four velocity of the fluid in comoving
coordinates. It is worthwhile to notice that in the presence of a
non-minimal  coupling between matter and curvature, the
energy-momentum tensor is not covariantly conserved which implies
that the motion of a point like particle is non-geodesic
[\refcite{Bertolami08a}].

Now, we define an effective energy-momentum tensor in equation (2)
thus the field equation recast in the form of the standard Einsteins
equation
\begin{equation}
G_{\mu\nu }=R_{\mu\nu }-\frac{1}{2}R g_{\mu\nu }=T_{\mu\nu
}^{(eff)},
\end{equation}
where $T_{\mu \nu }^{(eff)}$ is defined by
\begin{equation}
T_{\mu \nu }^{(eff)}=T_{\mu \nu }^{(m)}+T_{\mu \nu }^{(c)},
\end{equation}
where we have defined $T_{\mu \nu }^{(m)}$ and $T_{\mu\nu}^{(c)}$ as
follows

\begin{equation}
T_{\mu \nu }^{(m)} =\frac{1}{f_{1}^{'} +2\lambda f_{2}^{'}
{{\cal{L}}_{m}} } T_{\mu \nu },
\end{equation}

\begin{footnotesize}
\begin{equation}
T_{\mu\nu}^{(c)}=\frac{1}{f'_1+2\lambda
f'_2{{\cal{L}}_{m}}}\Big[\frac{1}{2}(f_1-f'_1R)g_{\mu\nu}-\lambda
f'_2R{{\cal{L}}_{m}}
g_{\mu\nu}+(\nabla_{\mu}\nabla_{\nu}-g_{\mu\nu}\Box)(f'_1+2\lambda
f'_2{{\cal{L}}_{m}})+\lambda f_2T_{\mu\nu}\Big]
\end{equation}
\end{footnotesize}
respectively. We assume that the matter Lagrangian density is
${{\cal{L}}_{m}}=-\rho_{m}$ ( for a discussion about the possibility
of other choices for the matter Lagrangian see for instance
Ref. \refcite{Bertolami08b} and references therein). Using the flat
Friedman-Robertson-Walker metric $ds^{2}=-dt^{2}+a^{2}(t)d{\bf
x}^{2}$, the equations of motion (2) lead to the modified Friedmann
equations as follows
\begin{equation}
3H^{2}=T_{tt}^{(m)}+T_{tt}^{(c)}\,,
\end{equation}
\begin{equation}
-\dot{a}^2-2 \ddot{a} a =T_{rr}^{(m)}+T_{rr}^{(c)}\,.
\end{equation}
where $H = \dot{a}/a$ is the  Hubble parameter. The latter equation
can be reduced to
\begin{equation}
-{\dot{a}^2 \over a^{2}}-{2 \ddot{a} a \over  a^{2}}=p_{m}+p_{c}
\end{equation}
where $p_{m}$ and $p_{c}$ are the pressure of the matter and
curvature fluid respectively. In the case of dust matter with
$p_{m}=0$ and regarding to the fact $R=6(2H^{2}+H\dot{H})$,
equations (10) and (11) can be rewritten as

$$
3H^2={R\over 2}-{f_1\over 2(f_{1,R}-2\lambda
f_{2,R}\rho_m)}-{3H^2f_{1,RR}R'\over (f_{1,R}-2\lambda
f_{2,R}\rho_m)}+{6H^2\lambda f_{2,RR}R'\rho_m\over (f_{1,R}-2\lambda
f_{2,R}\rho_m)}+ {6H^2\lambda f_{2,R}\rho'_m\over (f_{1,R}-2\lambda
f_{2,R}\rho_m)}$$
\begin{equation}
+{\lambda f_2\rho_m\over (f_{1,R}-2\lambda
f_{2,R}\rho_m)}+{\rho_m\over (f_{1,R}-2\lambda f_{2,R}\rho_m)}
\end{equation}
and
\begin{equation}
p_{c}=-\left(\frac{R}{3} -{H}^{2}\right)\,.
\end{equation}
respectively, where $f_{i,R}\equiv{\rm d}f_i/{\rm d}R$,
$f_{i,RR}\equiv{\rm d}^{2}f_i/{\rm d}R^{2}$ and $f_{i,RRR}\equiv{\rm
d}^{3}f,i/{\rm d}R^{3}$. Here we have defined
\begin{equation}
'=\frac{d}{d{\rm
 ln}a}\equiv\frac{d}{dN}=\frac{1}{H}\frac{d}{dt}
\end{equation}
and $\rho_m$ represent the matter energy density which is conserved
by virtue of the continuity equation via the relation
\begin{equation}
\dot{\rho}_{{\rm m}}+3H\rho_{{\rm m}}=0\,.
\end{equation}
Since a comprehensive study of the effect of a nontrivial $f_1(R)$
and $f_2(R)$ is complicated, in the rest of this paper, we take two
ansatzes. In the first approach, we set $f_1(R)=2R$ and
$f_2(R)=f(R)$ and in the second case we set
 $f_1(R)=f(R)$ and $f_2(R)=2R$ to study the dynamical behavior of the system.

\section{Autonomous Equations For the Case $f_1(R)=2R$ And $f_2(R)=f(R)$}
In the case $f_1(R)=2R$ and $f_2(R)=f(R)$, the Friedmann equations
(13) and (14) can be rewritten as
\begin{equation}
3H^{2} =\frac{1}{2-2\lambda
f_{,R}\rho_{m}}\rho_{m}+\frac{1}{2-2\lambda
f_{,R}\rho_{m}}\Big\{-\lambda f_{,R}R\rho_{m} +6\lambda
H^{2}f_{,RR}R'\rho_{m} -18\lambda H^{2}f_{,R}\rho_{m}+\lambda f
\rho_{m}\Big\}
\end{equation}
and
$$
-\left(\frac{R}{3} -{H}^{2} \right)=\frac{-\lambda }{2-2\lambda
f_{,R}\rho _{m}}\Big
\{2H^{2}R'^{2}f_{,RRR}\rho_{m}+2f_{,RR}HH'R'\rho_{m}+2
f_{,RR}H^{2}R''\rho_{m}+6H^{2}f_{RR}R'\rho_{m}-
$$
\begin{eqnarray}
f_{,R}R\rho_{m}-6HH'f_{,R}\rho_{m}-12f_{,RR}H^2R'\rho_{m} \Big\}
\end{eqnarray}
respectively. We can express the generalized friedmann equations
(17) and (18) in an autonomous system of the first ODEs to study the
cosmological dynamics of the model
\cite{Carloni05,Amendola07,Nozari12,Nozari13,Nesseris07}. For this purpose, we
express the generalized friedmann equation (17) in a dimensionless
form as follows {\footnotesize
\begin{equation}
1={1 \over 6}\frac{\rho_{m}}{H^{2}(1-\lambda
f_{,R}\rho_{m})}+\frac{\lambda f_{,RR}R'\rho_{m}}{1-\lambda
f_{,R}\rho_{m}}-{1 \over 6} \frac{\lambda
f_{,R}R\rho_{m}}{H^{2}(1-\lambda f_{,R}\rho_{m})} -3\frac{\lambda
f_{,R}\rho_{m}}{1-\lambda f_{,R}\rho_{m}}+{1 \over 6}\frac{\lambda
f\rho_{m}}{H^{2}(1-\lambda f_{,R}\rho_{m})}\,.
\end{equation}}
We define the dimensionless variables $x$, $y$, $z$ and $v$ as
\begin{equation}
x=\frac{\lambda f_{,RR}R'\rho_{m}}{1-\lambda f_{,R}\rho_{m}}\,,
\end{equation}
\begin{equation}
y=-{1 \over 6}\frac{\lambda f_{,R}R\rho_{m}}{H^{2}(1-\lambda
f_{,R}\rho_{m})}\,,
\end{equation}
\begin{equation}
z=-3\frac{\lambda f_{,R}\rho_{m}}{1-\lambda f_{,R}\rho_{m}}\,,
\end{equation}
and
\begin{equation}
v={1 \over 6}\frac{\lambda f\rho_{m}}{H^{2}(1-\lambda
f_{,R}\rho_{m})}
\end{equation}
respectively. By defining the density parameter $\Omega={1 \over
6}\frac{\rho_{m}}{H^{2}(1-\lambda f_{,R}\rho_{m})}$, the equation
(19) takes the form
\begin{equation}
\Omega={1 \over 6}\frac{\rho_{m}}{H^{2}(1-\lambda
f_{,R}\rho_{m})}=1-x-y-z-v\,.
\end{equation}
Using equ. (18), differentiating $x$ with respect to $N$ leads to
\begin{equation}
x'=-3{xy \over z}+6{y \over z}+x^{2}+2x-6y+2z+xz-1\,.
\end{equation}
Similarly, differentiating the equation (21) with respect to $N$
results
\begin{equation}
y'=-3\frac{xy}{mz}-3\frac{xy}{z}-6 {y^{2} \over z} +xy+yz+y\,,
\end{equation}
where
\begin{equation}
m=\frac{\ln F}{\ln R}=\frac{Rf_{,RR}}{f_{,R}}\,.
\end{equation}
Also, differentiating (22) and (23) leads to
\begin{equation}
z'=z^{2}+xz-3x-3z
\end{equation}
and
\begin{equation}
v'=3\frac{xy}{mz}-6\frac{yv}{z}+v+xv+zv
\end{equation}
respectively. The effective equation of state parameter is defined
as
\begin{equation}
w_{eff}=-1-\frac{2}{3} \frac{H'(N)}{H(N)}\,.
\end{equation}
using equation (15) and (18), we can rewrite this relation in the
following form
\begin{equation}
w_{eff}=-2 \frac{y}{z}+{1 \over 3}\,.
\end{equation}
Now our dynamical system depends on $m(f(R))$. We can eliminate $m$
from the system by solving Equ.(26) for ${xy \over zm}$ and
substituting in (29) with definition $s={y\over z}$, consequently
equation (29) can be rewritten as $v'=-sz'-3\frac{xy}{z}-6{y^{2}
\over z}+xy+yz+y-6\frac{yv}{z}+v+xv+zv$. If we substitute $z'$ from
(28),then our dynamical system is given by as follows
\begin{equation}
x'=-3sx+6s+2x-6sz+2z+x^{2}+xz-1\,,
\end{equation}
\begin{equation}
z'=z^{2}+xz-3x-3z\,,
\end{equation}
\begin{equation}
v'=4sz-6s^{2}z-6sv+v+xv+zv\,.
\end{equation}
In order to study the dynamics of the system of equations (32),(33)
and (34), we set $x'=0$, $z'=0$ and $v'=0$ and find the critical
points and their stability in each one of the cosmic eras. The
critical points of the scenario and the corresponding $\Omega$  are
summarized in table. \ref{ta1}. A general eigenvalues matrix is
obtained as below
\[ \left[ \begin{array}{c}
\frac{1}{2}\left(-1+3z+3x-3s+1\sqrt{1-2z-2x+42s+{z}^{2}+2xz-
18sz+{x}^{2}-6sx+9{s}^{2}}\right)\\
\frac{1}{2}\left[-1+3z+3x-3s-1\sqrt{1-2z-2x+42s+{z}^{2}+2xz-
18sz+{x}^{2}-6sx+9{s}^{2}}\right]\\
-6s+1+x+z  \end{array} \right]\]

\begin{table}[h]
\tbl{The critical points of the dynamical system (32), (33) and
(34).}
{\begin{tabular}
 {|c|c|c|c|c|c|}
\hline
point & $x$ & $y$ & $z$ & $v$& $\Omega$ \\
\hline \hline
A &${1-6s \over 3s}$ & $-{1-6s \over3}$& $-{1-6s \over 3s}$& $-2s+{4 \over 3}$&0 \\
\hline B &${3 \over 2}s-{5 \over 2}+{1 \over 2} \sqrt{9s^{2}+18s+5}$
&$3s$ & $3$ &${ 12s-18s^{2} \over {9 \over 2}s-{3 \over 2}-{1 \over
2} \sqrt{9s^{2}+18s+5}}$&$-{ 12s-18s^{2} \over {9 \over 2}s-
{3 \over 2}-{1 \over 2} \sqrt{9s^{2}+18s+5}}-{9 \over 2}s+{1\over 2}-{1 \over 2} \sqrt{9s^{2}+18s+5}$ \\
\hline C &${3 \over 2}s-{5 \over 2}-{1 \over 2}\sqrt{9s^{2}+18s+5} $
& $3s$ & $3$ & ${12s-18s^{2} \over {9 \over 2}s-{3 \over 2}+{1 \over
2} \sqrt{9s^{2}+18s+5}}$ &$-{ 12s-18s^{2} \over {9 \over 2}s-{3
\over 2}+{1 \over 2}
\sqrt{9s^{2}+18s+5}}-{9 \over 2}s+{1\over 2}+{1 \over 2} \sqrt{9s^{2}+18s+5}$ \\
\hline
\end{tabular}\label{ta1} }
\end{table}

The corresponding effective equation of state is
\begin{equation}
w_{eff}=-2s+{1 \over3}\,.
\end{equation}
 One can set favorable values for $w_{eff}$ which are related to the
several eras of the cosmic evolution and obtain the corresponding
$s$. We study the system in which the universe undergoes through the
radiation era ($w_{eff}=\frac{1}{3}$), matter era ($w_{eff}=0$) and
de Sitter era ($w_{eff}=-1$) to mimic a de Sitter late time
cosmology \cite{Nesseris07}.

\begin{figure}[h]
\epsfig{figure=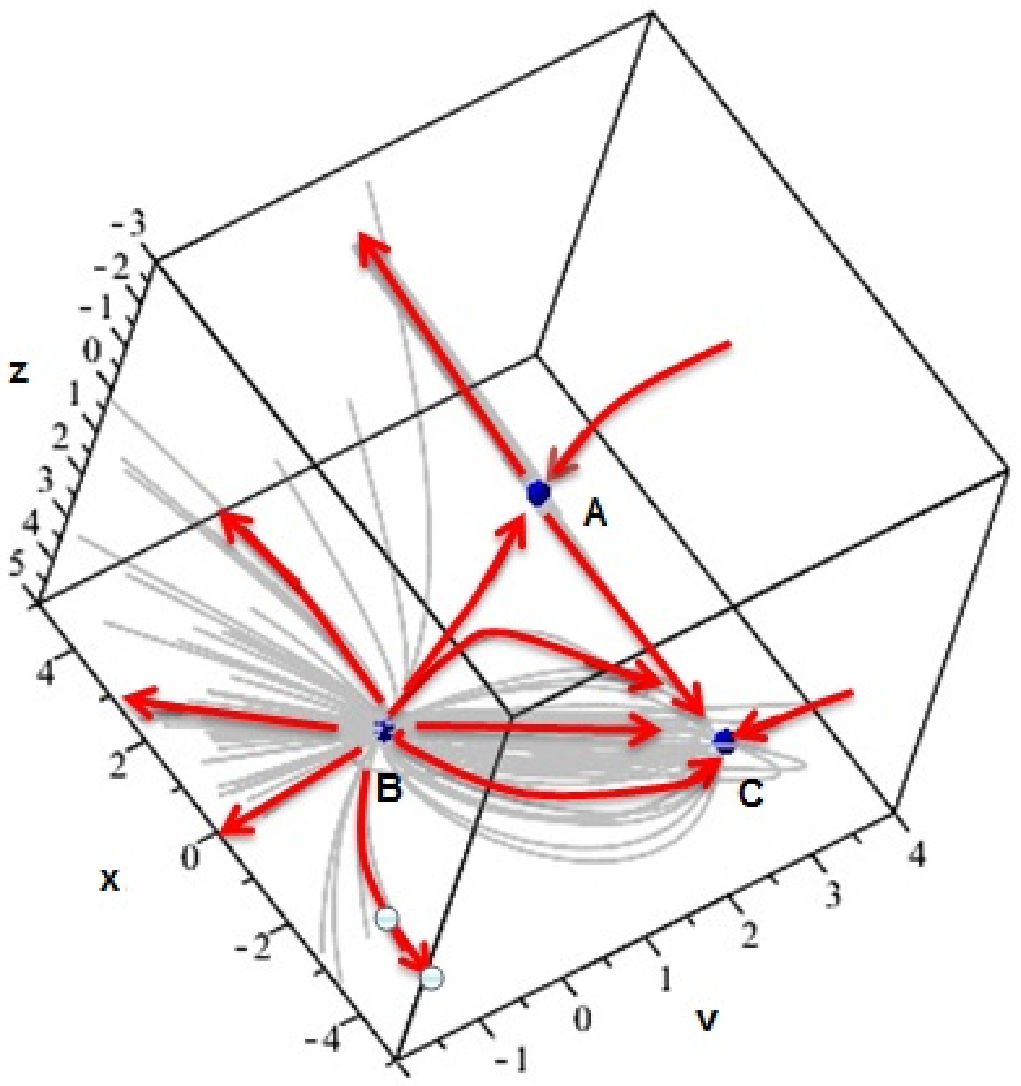,width=6.6cm}\epsfig{figure=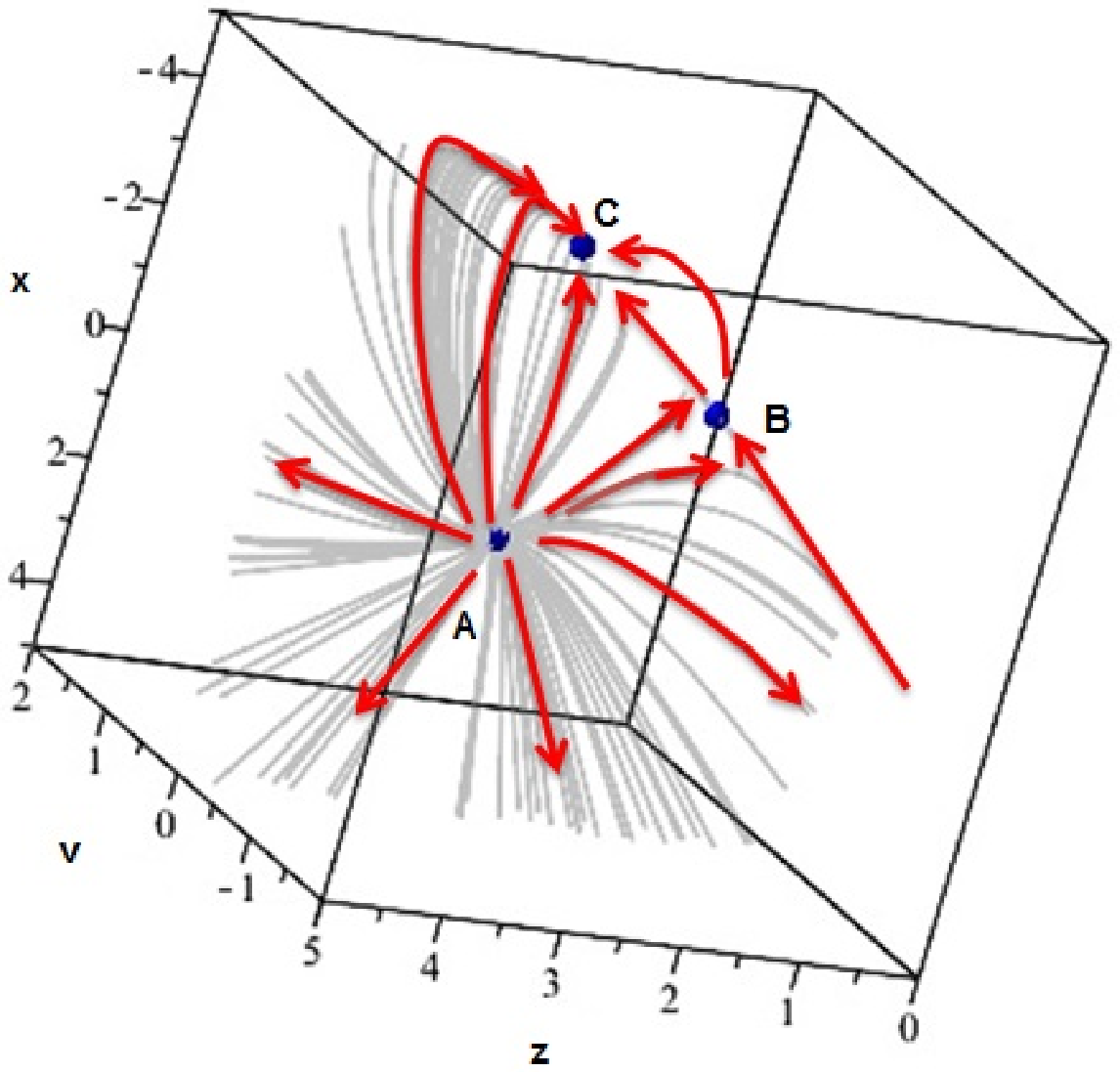,width=6.6cm}
\vspace*{6pt}\caption{The 3D phase space $x-v-z$ behavior for the
matter era (left panel) and the de Sitter era (right
panel).\protect\label{fig1}}
\end{figure}

The critical points in radiation, matter and de Sitter eras and
their stability in each eras are shown in table \ref{ta2}. It is
clear that for the radiation era, we have two critical points which
one of them is a saddle point and the other is unstable. For the
matter era, there exist three critical points which one of them is a
saddle point which corresponds to the usual matter era. In the de
Sitter era we also find three critical points which one of them is a
stable point corresponding to the late time acceleration of the
cosmic evolution. In Fig. \ref{fig1} (left panel), we depict the
corresponding $x-z-v$ phase space behavior for the matter dominated
era. As this figure shows, the critical point $B\,(-0.81,3,-0.68)$
is an unstable point and the point $C\,(-3.68,3,2.18)$ is a stable
point because the trajectories of the phase space are attracted by
this critical point. On the other hand, the point $A\,(0,0,1)$ is a
saddle point which is a suitable point to the matter dominated
phase. The $x-z-v$ phase space behavior for the de Sitter dominated
era is plotted in the right panel of the Fig. \ref{fig1}. This
figure shows that the critical point $B\,(-1.5,1.5,0)$ is a saddle
point and the phase space trajectories are passed from this point
and attracted by the stable critical point $C\,(-3.79,3,0)$. So the
critical point $C$ is a suitable point for the de sitter dominated
phase corresponding to the late time acceleration of the universe.
\begin{table}[h]
\tbl{The critical points of the system of equations (32),(33) and
(34) and their stability in each one of the three eras. Source
points (stable points) have only negative eigenvalues, saddle points
have mixed sign eigenvalues and sink points (unstable points) have
positive eigenvalues.} {\begin{tabular}{cccccccc}
\hline \hline \hline\\
\vspace{6pt}\textbf{Era}&\hspace{7pt}\textbf{$x$}\hspace
{7pt}&\hspace{7pt}\textbf{$y$}\hspace{7pt}&\hspace{7pt
}\textbf{$z$}\hspace{7pt}&\hspace{7pt} \textbf{$v$}
\hspace{7pt}&\hspace{7pt} \textbf{$\Omega$}
\hspace{7pt}&\hspace{7pt}
 \textbf{$w_{eff}$}\hspace{7pt}&\hspace{7pt}\textbf{Eigenvalues}\hspace{7pt}\\
\hline
\vspace{6pt} \textbf{Radiation} & -1.38 & 0 & 3 & 0 & -0.62 & $\frac{1}{3}$ & (2.62, 2.24, 1.62)\\
\vspace{6pt} \textbf{Radiation} & -3.61 & 0 & 3 & 0 & 1.61 & $\frac{1}{3}$ & (0.39, -0.61, -2.22)\\
\hline
\vspace{6pt} \textbf{Matter} & 0 & 0  & 0 & 1 & 0 & 0 & (0, 0.68, -2.1)\\
\vspace{6pt} \textbf{Matter} & -0.81 & 0.5 & 3 & -0.68 & -1 & 0 & (2.19, 2.88, 2.19)\\
\vspace{6pt} \textbf{Matter} & -3.68 & 0.5 & 3 & 2.18 & -1 & 0 & (-0.69, -0.69, -2.88)\\
\hline
\vspace{6pt} \textbf{de Sitter} & -1.5 & 1 & 1.5 & 0 & 0 & -1 & (0.79, -3.79, -3)\\
\vspace{6pt} \textbf{de Sitter} & 0.79 & 2 & 3 & 0 & -4.79 & -1 & (4.58, 3.79, 79)\\
\vspace{6pt} \textbf{de Sitter} & -3.79 & 2 & 3 & 0 & -0.21 & -1 & (-0.79, -4.58, -3.79)\\
\hline \hline \hline
\end{tabular}\label{ta2} }
\end{table}

\section{Autonomous Equations For the Case $f_1(R)=f(R)$ And $f_2(R)=2R$}
In this section, we allow that $f_1(R)$ to be an arbitrary function
of $R$ and set $f_2(R)=2R$, for little values of the coupling
parameter $\lambda$, the field equation (13) can be approximated as
\begin{equation}
1=-{f \over 6H^{2}f_R}-{2\lambda f \rho_m \over  6H^{2}f^2_R}+{R
\over 6H^{2}}-{2\lambda f_{RR} R'\rho_m \over f^2_R}-{f_{RR} R'
\over f_R}-{6\lambda \rho_m \over f_R}+ {\lambda R \rho_m \over 3H^2
f_R}+{2 \lambda \rho^2_m \over 3H^2 f^2_R} +\Omega_m
\end{equation}
where
\begin{equation}
\Omega_m={\rho_m\over 3H^2f_R}
\end{equation}
is the density parameter of the matter field and prime denote a
derivative with respect to $\ln a$. In the rest of this section, we
consider a power-law model for $f(R)$, i.e. $f(R)=R^n$, then the
generalized Friedman equation (36) takes the form
\begin{equation}
1=({n-1\over n}){R\over6H^2}+({n-1\over n^2}){2\lambda
\rho_m\over3H^2R^{n-2}} -(n-1){R'\over R}-{12\lambda \rho_m\over
nR^{n-1}}-4({n-1\over n}){R'\lambda \rho_m\over R^n}+{4\lambda
\rho^2_m\over3H^2 n^2 R^{2n-2}}+\Omega_m
\end{equation}
To study the phase space analysis of this scenario, we shall
introduce the following dimensionless variables
\begin{equation}
x_1=-(n-1){R'\over R}\,,
\end{equation}

\begin{equation}
x_2=-{12\lambda \rho_m\over nR^{n-1}}\,,
\end{equation}

\begin{equation}
x_3=({n-1\over n}){R\over6H^2}\,,
\end{equation}

\begin{equation}
x_4=({n-1\over n^2}){2\lambda \rho_m\over3H^2R^{n-2}}\,,
\end{equation}

\begin{equation}
x_5={4\lambda \rho^2_m\over3H^2 n^2 R^{2n-2}}\,,
\end{equation}

\begin{equation}
x_6=-4({n-1\over n}){R'\lambda \rho_m\over R^n}\,,
\end{equation}
Similarly to the previous section, the equations of motion for the
autonomous equations (39)-(44) are obtained as follows
{\footnotesize
$$
x'_1={3\over 3-x_2}\Big(x^2 -{n\over n-1}x_1x_3-{n\over n-1}
x_2x_3-{n\over n-1}x_3x_6+x_1x_6$$
\begin{equation}
-{n-3\over n-1}x_3+2x_2-x_1+{3 \over n-1}x_4-x_6-1\Big)\,,
\end{equation}

\begin{equation}
x'_2=-3x_2+x_1x_2\,,
\end{equation}
\begin{eqnarray}
x'_3=-{1\over n-1}x_1x_3-2{n\over n-1}x^2_3+4x_3\,,
\end{eqnarray}

\begin{eqnarray}
x'_4=-2{n\over n-1}x_4x_3+x_4-{1\over3}{n-2\over n-1}x_1x_2x_3\,,
\end{eqnarray}

\begin{equation}
x'_5=-2x_5-2{n\over n-1}x_3x_5+{2n-2\over n-1}x_5x_1\,
\end{equation}
and
$$
x'_6=-{x_2\over 3-x_2}(x^2 -{n\over n-1}x_1x_3-{n\over n-1}
x_2x_3-{n\over n-1}x_3x_6+x_1x_6$$
\begin{equation}
-{n-3\over n-1}x_3+2x_2-x_1+{3 \over n-1}x_4-x_6-1)-3x_6+x_1x_6\,.
\end{equation}
With the constraint equation (38), the density parameter of the
matter field can be obtained as $\Omega_m={\rho_m\over
3H^2nR^{n-1}}=1-x_1-x_2-x_3-x_4-x_5-x_6$. Also, with the definition
(30), the effective equation of state parameter is
$w_{eff}=-{2\over3}{n\over n-1}x_3+{1\over 3}$. In order to
investigate the dynamics which is implied by the equations of
motions, we set $x'_i=0$ (i=1-8) and find the critical points which
are shown in table 3.
 Where we have defined $A=\sqrt{84n^4-252n^3+253n^2-90n+9}$\,.\\
\begin{table}[h]
\tbl{The critical points of the dynamical system (45-50).}
{\begin{tabular}{l l l l l }\hline\hline Points&Coordinates
$(x_{1},x_{2},x_{3},x_{4},x_{5},x_{6})$
&$\Omega_m$&$w_{eff}$\\[0.5 ex]
\hline $P_{1}$&$(0,{1\over3},0,0,0,-{1\over3})$&$0$&${1\over
3}$\\[2 ex]
$P_{2}$&$(1.61,0,0,0,0,0)$&$-0.61$&${1\over 3}$\\[2 ex]
$P_{3}$&$(-0.61,0,0,0,0,0)$&$1.61$&${1\over 3}$\\[2 ex]
$P_{4}$&$(3,{-6n^2+30n-21\over 4n-7},{4n-7\over
2n},{2n^2-10n+7 \over 2n},0,{6n^2-30n+21\over 4n-7})$&$-n+1$&$\frac{2-n}{n-1}$\\[2 ex]
$P_{5}$&$\left(0,{3(n^2-3n+2)\over n^2-3n+4},{2(n-1)\over
n},{2(-n^3+5n^2-8n+4)\over n(3n^2-5n+4)},0,{3(-n^2+3n-2)\over
3n^2-5n+4}\right)$&${-n^2+n+2\over
3n^2-5n+4}$&$-1$\\[2 ex]
$P_{6}$&$(3n-3,0,{1\over 2}{n-1 \over n},{1\over
2}{-23n^2+21n^3-6n^4+7n+1\over n},0,0)$&${17\over 2}n-{21\over
2}n^2+3n^3$&
$0$\\[2 ex]
$P_{7}$&$\left(\frac{6n^2-7n+3-A}{2n(2n-1)},0,\frac{16n^3-30n^2+15n-3+
A}{4n^2(2n-1)},0,0,0\right)$&$\frac{-10
n^2+15n-3+A}{4n^2}$&$\frac{24n^2-12
n^3-13n+3-A}{6n(n-1)(2n+1)}$\\[2 ex]
$P_{8}$&$\left(\frac{6n^2-7n+3+A}{2n(2n-1)},0,\frac{16n^3-30n^2+15n-3-
A}{4n^2(2n-1)},0,0,0\right)$&$\frac{-10
n^2+15n-3-A}{4n^2}$&$\frac{24n^2-12
n^3-13n+3+A}{6n(n-1)(2n+1)}$\\[2 ex]
 \hline\hline
\end{tabular}\label{ta3} }
\end{table}
Among the critical points of the dynamical system, we discuss some
important cases of them as follows
\begin{itemize}\label{item1}
\item The Point $P_4$\\ This point for $n \to 1^+ $ and $n \to 1^- $ has
$\Omega_{de}\approx 1$ with $w_{eff} \to +\infty $ and $w_{eff} \to
-\infty $ respectively which are not cosmologically acceptable. The
point $P_4$ for $n=0$ has $\Omega_m=1$ but some of the critical
points disappear for that so it
 may be intersting to consider the scenario with $n \to 0^+$ and $n \to 0^-$ which
  both have $w_{eff} \approx -2$ which are unacceptable too.
\end{itemize}
\begin{itemize}\label{item2}
\item The Point $P_5$\\ This point is, independent of the value of $n$, represent a de
Sitter phase ($w_{eff}=-1$). So, we expect that the energy density
fraction of the matter must be zero ,i.e, $\Omega_m=0$. To have a
dark energy dominated era with $\Omega_m=0$ and $\Omega_{de}=1$, the
values of $n$ should be $n=2$ and $n=-1$ (see Fig. \ref{fig2}),
which leads to the critical points $(0,0,1,0,0,0)$ and
$(0,\frac{3}{2},4,-3,0,-\frac{3}{2})$ respectively. The
corresponding eigenvalues of these critical points are spiral
stables.
\end{itemize}
\begin{itemize}\label{item3}
\item The Point $P_6$\\ Since this point has an energy density fraction of the matter, this
 solution can be regarded as a scaling solution. This point is, independent
 of the value of $n$, shows a matter dominated phase with $w_{eff}=0$ and we expect that some $n$
leads to $\Omega_m=1$. Indeed for $n=1$, $n=0.143$ and $n=2.358$
(see Fig. \ref{fig2}), we have $\Omega_m=1$. In the case $n=1$, the
eigenvalues matrix diverges . However, if $n\rightarrow1^{+}$ or
$n\rightarrow1^{-}$, this point represents a saddle matter epoch
with oscillation. The values $n=0.14$ and $n=2.358$ leads to a
saddle matter epoch with oscillation too.
\end{itemize}

\begin{itemize}\label{item4}
\item The Point $P_7$\\In the case $n=2$, this point is corresponding to the de Sitter
point P5. Case $n=1$, gives rise to the matter dominated era (i.e.,
$\Omega_m=1$ and $w_{eff}=0$), but its eigenvalues become singular
and we cannot explain its stability. However, if
$n\rightarrow1^{+}$, there exists an unstable matter era with
$w_{eff}=0^-$, but if $n\rightarrow1^{-}$ leads to a saddle
oscillating matter epoch $w_{eff}=0^+$.
\end{itemize}
\begin{figure}[h]
\epsfig{figure=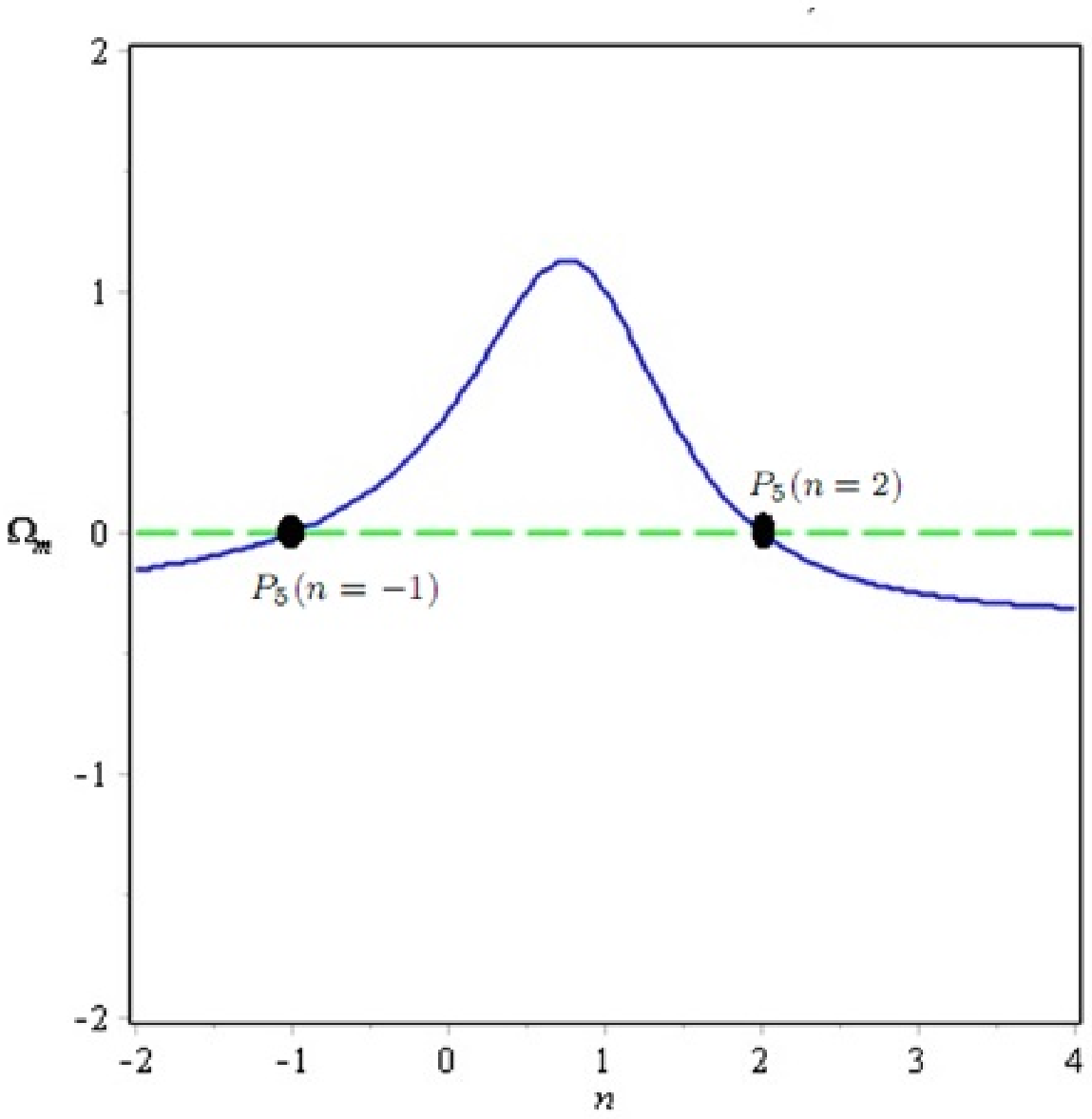,width=6.6cm}\epsfig{figure=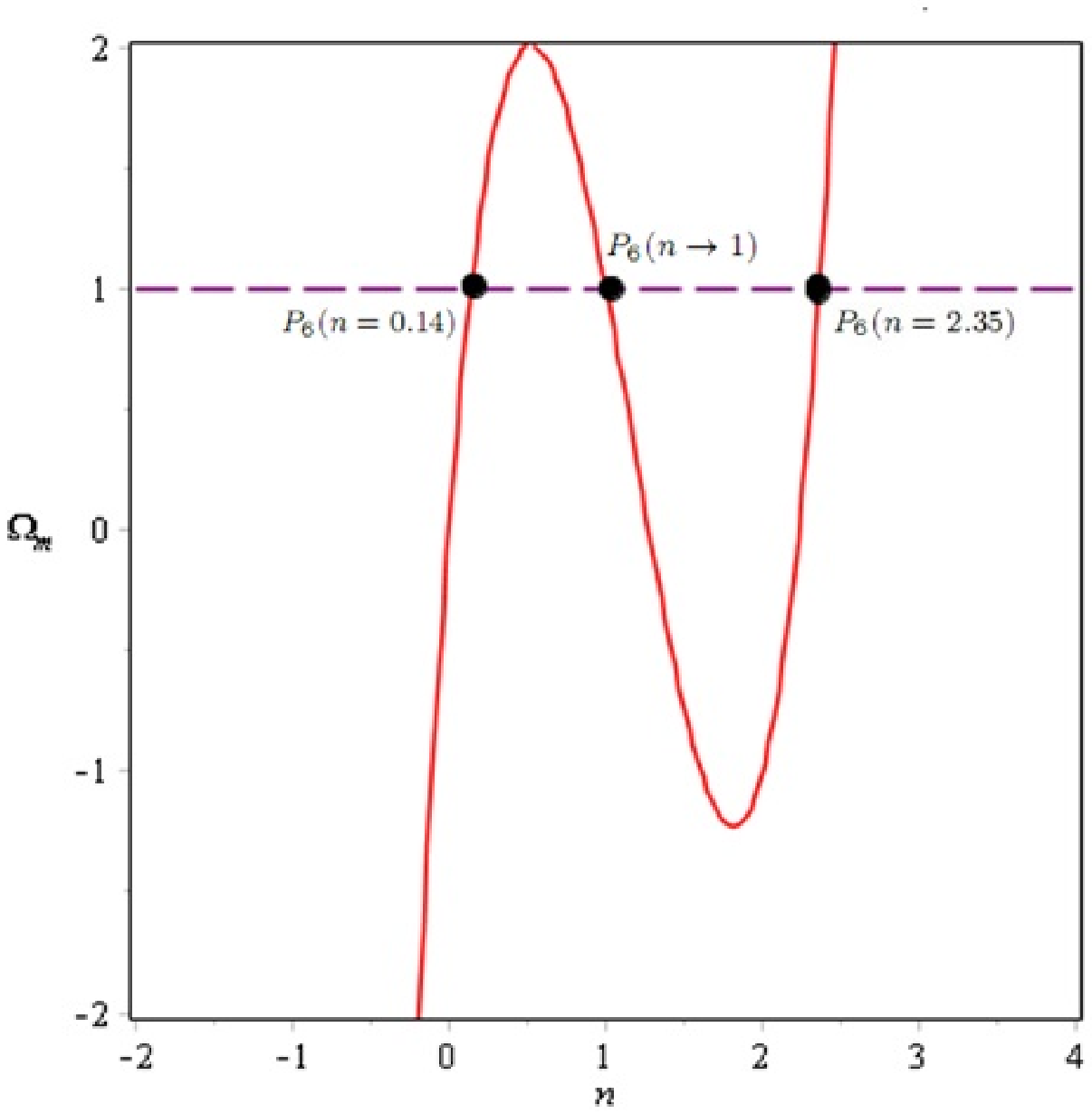,width=6.6cm}
\vspace*{6pt}\caption{(left panel) The plot of $\Omega_m$ versus $n$
for the point $P_5$. For $n=-1$ and $n=2$, the matter density
parameter vanishes which is corresponding to a de Sitter dominated
phase. (right panel) behavior of $\Omega_m$ for the point $P_6$. As
the figure shows, for $n=1, 0.1, 2.35$ the value of the matter
density parameter is equal to $1$ which corresponds to a matter
dominated phase.\protect\label{fig2}}
\end{figure}

To have a viable cosmological model, it has to possess a matter
dominated era prior to a stable accelerated expansion epoch. Now we
investigate the conditions that this transition can occur from
matter era to dark energy era. Point $P_6$ with $n=1$, $n=0.14$ and
$n=2.35$ is a good candidate for be in matter era because this point
 is a saddle oscillation point and ,once reached, it will give away
to a late-time acceleration. On the other hand, the effective
equation of state parameter and matter density parameter for this
point are $w_{eff}=0$ and $\Omega_m=1$ respectively which are
compatible to a matter era. Point $P_5$ for $n=2$ and $n=-1$ is a
stable point with $w_{eff}=-1$ and $\Omega_{de}=1$ therefore it is a
good choice for late-time attractor. So, we study the transition
from  point $P_6$ to $P_5$. If the points $P_5$ and $P_6$ had a
common value of $n$, we could say surely that the transition is
possible or not and the model for that $n$ is cosmologically viable.
But now we take one step forward and study the transition between
$P_6$ and $P_5$ for different values of $n$. We have five
possibilities of transition between these two points which study
them below:\\
\begin{romanlist}[(II)]
\item \, $P_6(n=2)\to P_5(n=2)$\\
In the case $n=2$, the point $P_5$, as studied above, is acceptable
to be a late-time attractor. On the other hand, the point $P_6$ has
$w_{eff}=0$ which is ideal for the matter point but the matter
density parameter is $-1$  which is cosmologically unacceptable.

\item \, $P_6(n=-1)\to P_5(n=-1)$\\
In this case, $P_5(n=-1)$ is acceptable to be a late-time attractor
but the point $P_6(n=-1)$ corresponds to $w_{eff}=0$ and
$\Omega_m=-22$,. So from cosmological point of view this case is not
acceptable. consequently, this transition is impossible.

\item \, $P_6(n \to 1)\to P_5(n \to 1)$\\
For $P_5(n \to 1)$, although the effective equation of state
parameter is corresponding to a de Sitter acceleration with
$w_{eff}=-1$ but in this case, this solution has $\Omega_{de} \to 0$
which is a "dark energy era" without dark energy, that is clearly
unacceptable.

\item \, $P_6(n=0.14)\to P_5(n=0.14)$\\
The case $P_5(n=0.14)$ has $\Omega_{de}<1$ which is not consistent
with the dark energy era as is studied above. So this transition is
not acceptable too.

\item \, $P_6(n=2.35)\to P_5(n=2.35)$
\\
In the case $n=2.35$, $P_6$ is a saddle with oscillation point with
$w_{eff}=0$ and $\Omega_m=1$ which is a good candidate to be in the
matter era. $P_5$ has $w_{eff}=-1$ and $\Omega_{de}=1.13$ which is
stable spiral point with eigenvalues
\[ \left[ \begin{array}{c}
-6\\-2.95\\-0.13\\-3.04\\-3.5+0.5i\\-3.5-0.5i \end{array} \right]\]
Since $P_5$ has $\Omega_{de}\approx 1$ with $w_{eff}=-1$ and is a
stable spiral point, it seems that the transition $P_6(n=2.35)$ to
$P_5(n=2.35)$ is acceptable. This issue is shown in Fig. \ref{fig3}.
\end{romanlist}
\begin{figure}[h]
\epsfig{figure=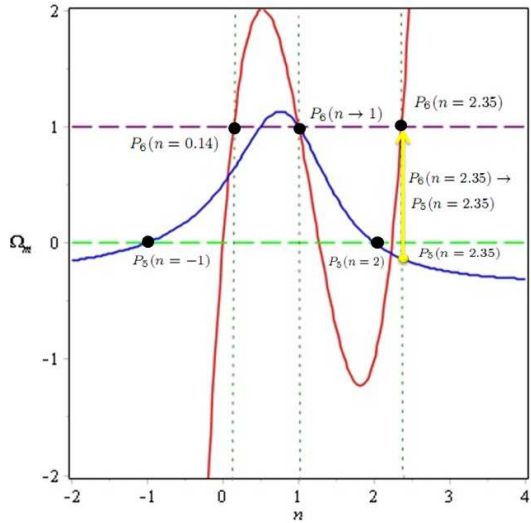, width=8cm} \vspace*{8pt}\caption{
The matter density parameter versus $n$ for the critical point $P_5$
(blue solid line) and the point $P_6$ (red solid line). Intersection
of the $P_5$ curve to the purple dashed line are corresponding to
the matter dominated phase and intersection of the $P_6$ curve to
the green dashed line are the de Sitter dominated phase.
\protect\label{fig3}}
\end{figure}
\section{Conclusion}
In this work, we have considered a generalized $f(R)$ gravity theory
with a non-minimal coupling between matter and curvature scalar and
studied the cosmological solution in a spatially flat FLRW
background. Since in this scenario we have two arbitrary functions
of Ricci scalar which one of them is coupled non-minimally to the
matter, the resulted equations of motion are very complicated. So to
perform a phase space analysis of this model, we have considered two
particular cases and in both of them can sequently we have set one
function equal to the Ricci scalar and allowed the other to be a
functional form of $R$. To study the dynamics implied by the
modified Friedmann equations, we have written the first Friedmann
equation in a dimensional form and obtained the autonomous system of
the first ODEs in each case. In the first case, we have assumed
$f_1(R)=2R$ and we have set favorable values for $w_{eff}$ (the
effective equation of state parameter) which is related to the
several eras of the cosmic evolution and find the corresponding
critical points. We have studied the system in which the universe
goes through the radiation era ($w_{eff}=\frac{1}{3}$), matter era
($w_{eff}=0$) and de Sitter era ($w_{eff}=-1$) to mimic the
$\Lambda$CDM cosmology. We have found that for the radiation era,
there are two critical points which one of them is a saddle point
and the other is unstable. For the matter era, there exist three
critical points which one of them is a saddle point which
corresponds to the usual matter era. In the de Sitter era we also
have found three critical points which one of them is a stable point
corresponding to the late time acceleration of the cosmic evolution.
In the second case, we have set $f_1(R)=f(R)$ and assumed a small
non-minimal coupling between the Ricci scalar and matter Lagrangian
density by a small value of the coupling parameter $\lambda$. To
investigate the cosmological dynamics in this case, we have assumed
a power law form for $f(R)$ and found the critical points in a
different manner to the previous case. By studying the stability
conditions of these critical points, we have found that in the phase
space of this model there are some saddle points and stable points
which are corresponding to the matter era and late time acceleration
era respectively. We also have investigated the conditions that the
transition can occur from the matter era to dark energy era and have
shown that depending on the some values of the power $n$, this model
has a saddle matter dominated era prior to a stable accelerated
expansion epoch.
\section*{Acknowledgements}
We are grateful to Prof. Kourosh Nozari for helpful discussions and
suggestions.
\providecommand{\newblock}{}

\end{document}